\documentstyle[11pt]{article}

\newcommand{\beq}{\begin{equation}}
\newcommand{\eeq}{\end{equation}}
\newcommand{\bea}{\begin{array}}
\newcommand{\ea}{\end{array}}
\newcommand{\beqa}{\begin{eqnarray}}
\newcommand{\eeqa}{\end{eqnarray}}
\newcommand{\bean}{\begin{eqnarray*}}
\newcommand{\eean}{\end{eqnarray*}}
\newcommand{\eqn}[1]{(\ref{#1})}

\newcommand{\del}{\partial}

\newcommand{\js}{Jordan--Schwinger\ }

\newcommand{\C}{\:\mbox{\sf C} \hspace{-0.82em} \mbox{\sf C}\,}
\newcommand{\R}{\mbox{I \hspace{-0.82em} R}}

\newcommand{\g}{\cal G}
\newcommand{\gi}{\tilde {\cal G}}

\def\up#1{\leavevmode \raise.16ex\hbox{#1}}

\newcommand{\journal}[4]{{\sl #1 }{\bf #2} \up(19#3\up) #4}

\newcommand{\gapproxeq}{\lower .7ex\hbox{$\;\stackrel{\textstyle >}{\sim}\;$}}
\newcommand{\lapproxeq}{\lower .7ex\hbox{$\;\stackrel{\textstyle <}{\sim}\;$}}

\newcounter{appendice}

\def\thebibliography#1{\section*{REFERENCES\markboth
 {REFERENCES}{REFERENCES}}\list
 {[\arabic{enumi}]}{\settowidth\labelwidth{[#1]}\leftmargin\labelwidth
 \advance\leftmargin\labelsep
 \usecounter{enumi}}
 \def\newblock{\hskip .11em plus .33em minus -.07em}
 \sloppy
 \sfcode`\.=1000\relax}

\begin{document}
\hfill INFN-NA-IV-93/38,DSF-T-93/38
\begin{center}
\vskip 0.8cm
{ \LARGE \bf  A generalization of the \js map: classical version and
 its q--deformation.}
\vskip 0.8cm
{V.I.Man'ko \footnote{on leave from Lebedev Physics Institute, Moscow},
G.Marmo,  P. Vitale and F.Zaccaria\\ Dipartimento di Scienze Fisiche\\
Universit\`a di Napoli "Federico II" \\ and\\Istituto Nazionale
 di Fisica Nucleare\\ Sezione di Napoli}\\Mostra d'Oltremare, Pad.19 - 80125
Napoli, Italy
\vskip 0.2cm
\end{center}

\vskip 0.8cm

\begin{abstract}
For all three--dimensional Lie algebras the construction
of generators in terms of functions on  4-dimensional real phase space
is given with a realization of the Lie product in terms of Poisson
brackets.
This is the classical Jordan--Schwinger map which is also given
for the deformed algebras ${\cal {SL}}_{q}(2,\R)$,
${\cal E}_{q} (2)$ and ${\cal H}_q(1)$. The ${\cal U}_{q}(n)$ algebra is
discussed in the same context.
\end{abstract}

\setcounter{footnote}{1}
\section{Introduction}
The Jordan-Schwinger map \cite{jord},\cite{schw}, is well known to
physicists in connection with the realization of the ${\cal {SU}}(2)$ algebra
in terms of polynomials of creation and annihilation operators,$a^{\dag}, a$,
for
 the two--dimensional harmonic oscillator. This map allows to construct all the
irreducible representations of the group $SU(2)$ using the states of
two--dimensional harmonic oscillator and the technique of creation and
annihilation operators: in fact  through this map the angular momentum
operators may be expressed as quadratic forms in $a_i^{\dag}, a_i$.
This map is a particular case of the second
quantization procedure due to which generators of any Lie algebra may be
realized as quadratic forms in creation and annihilation operators, of
bosonic or fermionic kind, if a matrix representation of this Lie algebra is
given.

This map has also been used (\cite{bied}, \cite{mcfar}) in
connection with realizations of ${\cal {SL}}_q(2)$, where generators may be
realized as quadratic forms of q-oscillators. Here and in the following we
shall mean with ${\cal {SL}}(2)$ the complex Lie algebra $A_1$ in the Cartan
classification, whose real forms are ${\cal {SU}}(2)$, ${\cal {SL}} (2, \R)$
and ${\cal {SU}}(1,1)$. When needed we will choose one of them.

An abstract Lie algebra, $\g$, may be realized not only in terms of
operators but also in terms of functions on Poisson manifolds (\cite{lich}),
the Lie bracket now being the Poisson bracket. Realizations of this kind
exist, in fact, naturally when Poisson manifolds are submanifolds of
$\g^*$, the dual of $\g$ (\cite{kks1},\cite{kks2},\cite{kks3}). It is then
natural to ask
whether Poisson maps exist between generic manifolds and $\g^*$.
In {\bf section 2} we will be
interested in finding these Poisson maps between $\C^2$, equipped with the
standard symplectic structure, and ${\cal {SL}} (2)^*$.
This is what we may call a classical \js map and it is already available in
the literature (\cite{flat}, \cite{frons}). As we already mentioned, it is
through the \js map that a realization of ${\cal {SL}}_q(2)$ was given in
terms of 2--dimensional q--oscillators; analogous objects can be
introduced in the classical setting. These classical hamiltonian systems may
be of interest {\it per se}, showing a particular non linearity in the
energy dependence of the frequency (\cite{mmsz}). Another reason for such a
study is that the two kinds of realizations are connected by the Dirac map
(\cite{cmp}), that is by canonical quantization procedure. This would allow
hopes to transfer results obtained in the classical context to the quantum
one, a part from ordering problems.

In {\bf section 3} all the three--dimensional Lie algebras are considered with
respect to their realization as Poisson algebras and the relevant maps,
which we will call generalized classical \js maps, are given.

The subject of {\bf section 4} deals with effects of deformation on the \js
map, while in {\bf sections 5} and {\bf 6} we comment on the
possibility to extend
the above constructions to higher dimensional Lie algebras, taking into
particular consideration ${\cal {SU}}_q(n)$.
%

%

\section{\bf The \js map and its classical analogue}

We review very briefly the \js  map as it was introduced by Schwinger
to deal with angular momentum in terms of creation and annihilation
operators.

A realization of the Lie algebra ${\cal {SL}}(2)$ in terms of creation and
annihilation operators, $a_i, a^{\dag}_i,~~i \in \{1,2\}$, satisfying standard
commutation relations
$$
[a_i, a_i^{\dag}]=\delta_{ij}
$$
is provided by
\beq
X_{+} = a_{1}^{\dag}a_{2}~,~~~~~~~~ X_{-} =
a_{2}^{\dag}a_{1}~,~~~~~~~~X_{3} =
\frac{1}{2}(\hat n_{1} - \hat n_{2})
\eeq
with $\hat n_{i} = a_{i}^{\dag}a_{i},  i = 1, 2$.
We have in fact
\beq
[X_{\pm}, X_{3}] = \mp X_{\pm}~~~~~~[X_{+}, X_{-}] = 2 X_{3}
. \label{su2com}
\eeq

To consider a classical analogue of this realization, we apply a
"dequantization procedure", i.e. we replace operators with functions on phase
space and commutators with Poisson brackets. To be definite we will consider
the Lie algebra of $SL(2, \R)$.
Usually Poisson brackets are defined in real phase space for
real-valued functions. As in our "dequantization
procedure" we replace creation and annihilation operators
by complex-valued functions on real
phase space, we have to extend the usual definition of Poisson brackets
to the case of complex-valued functions. We recall that this extension can be
defined by linearity.
For two complex-valued functions $f=f_{1}+if_{2}$ and $g=g_{1}+ig_{2}$ for
which
real and imaginary terms are real functions on the
phase space, we have
\beq
\{f,g\}=\{f_{1},g_{1}\}-\{f_{2},g_{2}\}+i(\{f_{1},g_{2}\}+\{f_{2},g_{1}\}).
\label{compbr}
\eeq
Here $\{f_{i},g_{k}\}$ are given by the standard Poisson brackets for real
functions on the phase space with coordinates $p_i, q_i$
\beq
\{p_{i},q_{j}\}=\delta _{ij},~~~~~~\{p_{i},p_{j}\}=\{q_{i},q_{j}\}=0.
\label{chc}
\eeq
Then for complex variables $z_i$,
$z_{k}^*$, $z_i = \frac{1}{\sqrt 2}(p_i+iq_i)$ the symplectic Poisson brackets
are given by
\beq
\{z_i^*, z_j\}=i\delta_{ij}~~.\label{pb}
\eeq
 A realization of $SL(2, \R)$ in terms of complex-valued functions
is obtained by setting
\beq
{\cal X}_{+} = i z_{1}^{*} z_{2} ,~~~~~~~~
{\cal X}_{-} = i z_{2}^{*} z_{1},~~~~~~~~
{\cal X}_{3} = \frac{i}{2}(z_1 z_1^* - z_2 z_2^*) \label{al}
\eeq
and we get
\beq
\{ {\cal X}_{\pm}, {\cal X}_{3}\} = \mp {\cal X}_{\pm},~~~~~~~
\{ {\cal X}_{+}, {\cal X}_{-}\} = 2 {\cal X}_{3},~~~~~~~\label{unre}
\eeq
This is what we mean by a classical realization of the Lie algebra ${\cal
{SL}}(2, \R)$, or the classical \js map for such an algebra.

To construct the angular momentum algebra in terms
of Poisson brackets, we have used  complex valued functions \eqn{al}.
The use of complex valued functions is convenient but
not mandatory. In fact the same algebra can be realized using
real functions of $q_i$, $p_i$, $i=1,2$.

The real analogue of \js map relations \eqn{al} are
\beq
{\cal Y}_{+}= p_{2}q_{1},~~~~~~{\cal Y}_{-}=
p_{1}q_{2},~~~~~~~
{\cal Y}_{3}=\frac {1}{2}(p_{1} q_1 - p_2 q_2)~.   \label{real}
\eeq
In fact, using the brackets \eqn{chc}, it can be checked that these functions
reproduce relations \eqn{unre}, i. e. they close on the algebra of $SL(2, \R)$.

We have shown therefore that it is possible to have Poisson brackets
realization of the Lie algebra commutation relations of $SL(2, \R)$ in two
ways.
One way uses the functions on $\R^4$ with values in $\R$ and the standard
Poisson brackets. The other one uses the extended definition of Poisson
brackets for complex valued functions on $\R^4$.

Let us now investigate the geometrical and algebraic meaning of what has been
done. We started with the
symplectic Poisson
algebra ${\cal F}(\R^4;\R)$  with the
standard Poisson bracket $\{p_a, q_b\}=\delta_{ab}$ and considered a
Poisson subalgebra generated by ${\cal Y}_{+},~{\cal Y}_{-},~{\cal Y}_{3}$.
These functions close on the Lie algebra of $SL(2, \R)$, say $\g$. They are
functions on $\R^4$ and linear functions on $\g^*$, the dual of $\g$,
thus they may be taken
as coordinates for $\g^*$. To put it more explicitly, we have constructed
a map from $\R^4$ to $\R^3 \equiv \g^*$ given by
$$
\pi:(p_1, q_1, p_2, q_2)\rightarrow ( p_{2}q_{1},~
p_{1}q_{2},~
\frac {1}{2}(p_{1}q_1 - p_2 q_2))
$$
With respect to the natural Poisson bracket on $\g^*$ this map is a Poisson
map, that is, identifying
$
{}~~~(p_{2}q_{1},~
p_{1}q_{2},~
\frac {1}{2}(p_{1} q_1 + p_2 q_2))~~~ $
with
$~~~(x_+, x_-, x_3) \in \g^*
$
and
$$
\pi ^* x^+ = {\cal Y}^+ \in {\cal F}(\R^4),~~~~
\pi ^* x^- = {\cal Y}^- \in {\cal F}(\R^4),~~~~
\pi ^* x^3 = {\cal Y}^3 \in {\cal F}(\R^4).
$$
we have
\beq
\{\pi^*(x_{\pm}),\pi^*(x_3)\}_{\R^4}=\pi^*(\{x_{\pm},x_{3}\}_{\g^*})
\eeq
and similarly for the other bracket.
We could use this observation to generalize the notion of \js map. We consider
a symplectic Poisson algebra ${\cal F}$ on a symplectic manifold $M$ and a
generic n-dimensional Lie algebra, so that our definition of (classical) \js
map would be:

{\bf Definition} ~~{\sl A classical generalized \js map for a
n--dimensional
Lie algebra $\g$ with
structure constants $c_{ij}^k$, is a
realization of $\g$ as a n--dimensional Lie subalgebra $\gi $ of
${\cal F}(M)$,
where ${\cal F}(M)$ is equipped with the standard Poisson bracket.}

\noindent If $\gi$ is realized by $f_1,...f_n$ with Poisson brackets
\beq
\{f_i, f_j\}=c_{ij}^k f_k \label{poisub}
\eeq
we may consider the Poisson algebra on ${\cal F}(\g^*)$ given by $\{x_i, x_j\}=
c_{ij}^k x_k$. We have an associated map  $\pi : M \rightarrow \g^*$,
characterized by $\pi^* x_i= f_i$.
We notice that this map needs not to be a projection, we will give indeed  some
 examples where $dim M < dim \g^* $.

Let us consider $\R^2 \equiv T^* \R$ with standard coordinates and Poisson
brackets
$$\{p,q\}=1.$$

1)$$\pi_1 : \R^2  \rightarrow \R^3$$
$$(p,q) \rightarrow (pq, \frac{q^2}{2}, -\frac{p^2}{2})
\equiv
({\cal H}, {\cal X}_+, {\cal X}_-) $$
and find
$$
\{ {\cal H}, {\cal X}_{\pm}\} = \pm 2 {\cal X}_{\pm}~~~~
\{ {\cal X}_+, {\cal X}_-\} = {\cal H}
$$
that is, we get a classical realization of the $SL(2,\R)$ algebra as a
subalgebra of ${\cal F}(\R^2)$.

2)$$\pi_2 : \R^2  \rightarrow \R^3$$
$$(p,q) \rightarrow (e^p, e^{-p}, q) \equiv ({\cal X},
{\cal Y}, {\cal Z}) $$
and find
$$
\{ {\cal X}, {\cal Z} \}={\cal X},~~~~\{ {\cal Y} , {\cal Z} \}=-{\cal Y},~~~~
\{ {\cal X},{\cal Y}\}=0
$$
which is the Poincar\'e algebra in three dimensions.

3) $$\pi_3 : \R^2  \rightarrow \R^3$$
$$(p,q) \rightarrow (sin~ q, cos~ q, p) \equiv ({\cal A},
{\cal B}, {\cal C}) $$
 and find
$$
\{ {\cal A}, {\cal C} \}=-{\cal B},~~~~\{ {\cal B} , {\cal C} \}={\cal A},~~~~
\{ {\cal A},{\cal B}\}=0
$$
which is the algebra of $E(2)$, the euclidean group in three dimensions.

Going back to the general situation, in order to give a basis
independent version of \eqn{poisub} let us consider
a $\R$-linear map on the second argument
\beq
F: M\times \g \rightarrow \R
\eeq
then \eqn{poisub} may be written in the following way
\beq
\{F(u), F(v)\}=F([u,v])  \label{2poisub}
\eeq
where $u,v\in \g$ and $F(u):M \rightarrow \R $.
If $u$ and $v$ are taken to be the generators of the Lie algebra, say
$e_i,~e_j$, then
\beq
\{F(e_i), F(e_j)\}=F([e_i,e_j])=c_{ij}^k F(e_k)
\eeq
which is equivalent to \eqn{poisub}.

Let us define
$$\hat F \equiv F_{\g} : M \rightarrow \g^* (\equiv Lin (\g , \R)).
$$
The map $\hat F$ could be identified with our previous definition of
the classical generalized \js map for a generic Lie algebra.

A third interpretation of the classical generalized \js map
is suggested by the last
map $\hat F$. Indeed
let us consider a symplectic manifold $(M, \omega)$ together with a Lie algebra
 $\g$ realized in terms of vector fields on $M$, say $X_1,...,X_n$,
\beq
[X_i, X_j] = c_{ij}^k X_k. \label{relX}
\eeq
 Consider
then the hamiltonian functions, $f_1,...,f_n$, associated with these vector
fields, namely
$$
i_{X_k} \omega = -df_k
$$
we have:
{\bf Proposition}~~ {\sl A classical generalized \js map is
equivalent to a strongly
symplectic action of $\g$ on $M$.}

\noindent We recall that an action is said to be strongly symplectic if
\eqn{relX} implies that
$$
\{ f_i, f_j \}= c_{ij}^k f_k
$$
 We would like to
notice that, to make our definition meaningful, the fields $X_i$
need not to be complete, i. e. it is not necessary that they integrate to a
group.

Previous examples show that when we consider a generalized \js map we should
not expect that the subalgebra will be realized in terms of quadratic maps
like for $SL(2, \R)$.

We can now ask if for any Lie algebra $\g$ there exists a realization in
terms of a Poisson subalgebra of a symplectic Poisson algebra ${\cal F}(M)$,
for some symplectic manifold M. The answer is positive, indeed if $\g$  is
the Lie algebra of a Lie group $G$, we can consider $M=T^*G$ and the
(left or right) momentum map $J:T^*G \rightarrow \g^*$ provides us with the
required realization. Moreover the momentum map is what we called $\hat F$.

It is possible to consider a symplectic manifold of smaller dimensions.
Indeed we can consider the Lie group $G$ and any maximal closed subgroup $K$ .
Here there is an action of $G$ on $G/K$, say the left action, if the quotient
was taken with respect to the right action of $K$. We set $M=T^{*}(G/K)$ and
the lifted action of $G$ to $T^{*}(G/K)$ is a strongly symplectic map
providing as with the required realization. In general, it is enough to
consider any manifold carrying a symplectic action of $G$ if $H^{2}(\g,\R)=0$,
$H^{1}(\g,\R)=0$.
Many examples are also constructed by starting with a representation of $G$,
i.e.
a linear action on some vector space $V$ and doubling the space by going to
$T^{*}V$ or $V^{\C}$. We shall consider this situation in Section 6.

We notice at this point that our "dequantization" and "quantization" procedures
associated with the Jordan-Schwinger map rely  on the correspondence
$a^{+}_{i}\leftrightarrow z^{*}_{i}, a_{i}\leftrightarrow z_{i}$.
Therefore, to be able to save this correspondence we shall try to find
a symplectic realization of our Lie algebras in terms of a standard structure
on $\R^{2n}$ or $\C^{n}$. We shall then say that we have a {\sl classical
Jordan-Schwinger map} (i.e. we drop the word 'generalized') if the Lie
algebra $\g$ can be realized as a Poisson sub-algebra of the real valued
or complex valued functions on $\R^{2n}$. To put it differently, we are
requiring that the symplectic manifold $(M, \omega)$ should be restricted to be
$(\R^{2n}, dp_{i}\wedge dq^{i})$.
We have decided to introduce the notion of generalized \js map to make
contact with the constructions associated with geometric quantization theory
(\cite{kks1}, \cite{kks2}, \cite{kks3}.
In the coming section we show that for any three-dimensional Lie algebra we
have a classical Jordan-Schwinger map on $T^{*}\R^{2}$.
We notice that, by using $T^{*}(G/K)$, with $K$ any closed one-dimensional
subgroup, it would be possible to give a generalized Jordan-Schwinger map
on a four-dimensional symplectic manifold. Our result will show that it is
possible to take our symplectic manifold to be $T^{*}\R^{2}$.

\section{Three dimensional algebras in terms of Poisson brackets}
As it is known (\cite{marpergrab}), Poisson structures on $\R^3$ can be
characterized by 1-forms
which admit an integrating factor. The argument goes along the following
lines. On $\R^3$ we consider coordinates $(x_1, x_2, x_3)$ along with a
volume form $\Omega = dx_1 \wedge dx_2 \wedge dx_3$. Any Poisson bracket
defines a bivector field
$$
\Lambda = c_{ij} (x) \frac {\del}{\del x_i} \wedge \frac {\del}{\del x_j}.
$$
By contracting $\Lambda$ with $\Omega$ we find
$$
i_\Lambda \Omega = \epsilon^{ijk} c_{ij}(x) dx_k
$$
i.e.
$$i_\Lambda \Omega = \alpha = A^k dx_k.
$$
 The Jacobi identity is equivalent to
$$
d\alpha \wedge \alpha = 0
$$
 i.e.
$$
\epsilon_{ijk} ( \frac {\del A^k}{\del x^i}-
  \frac {\del A^i}{\del x^k}) A^j =0.
$$

\noindent  Locally it means that
$\alpha
= f d\phi$ and $c_{ij}=\epsilon_{ijk} f \frac {\del \phi}{\del x_k}$.
Symplectic leaves for $\Lambda$, are characterized to be level sets of
$\phi$, say $\Sigma_a$,  or by the fact that the pull--back of $\alpha$ to
each leaf vanishes. The 1--form $\alpha$ will be called a Casimir 1--form
for the given brackets.

\noindent Now we are able to demonstrate

{\bf Theorem} {\sl Any 3--dimensional Lie algebra acts in a strongly symplectic
way on $(\R^4, \omega_0)$}.

The statement of the theorem is equivalent to say that every
three dimensional Lie algebra may be realized
as a Poisson subalgebra of ${\cal F}(\R^4)$ with the standard Poisson
structure, or which is the same, for any three--dimensional Lie algebra it
is possible to exhibit a classical \js map.

Let us demonstrate this explicitly.
The following Poisson brackets provide a realization of any real three
dimensional Lie algebra when the real parameters $a,~ b,~ c,~ n,$ are selected
accordingly (\cite{marpergrab}):
\beq
\{x,y\}=cw+ny,~~~~
\{y,w\}=ax,~~~~
\{w,x\}=by-nw. \label{classbr}
\eeq
These Poisson brackets satisfy the Jacobi
identity if and only if
\beq
na=0
\eeq
The Casimir 1--form is
\beq
\alpha=(cw+ny) dw + ax dx + (by -nw) dy
\eeq
or in more convenient form
\beq
\alpha = n(y dw-w dy) + d \frac{1}{2} (ax^2+by^2+cw^2)  \label{alph}
\eeq
with
\beq
d\alpha\wedge \alpha = n~a~x~ dy\wedge dw\wedge dx
\eeq
\noindent We will show by explicit construction that all the
three--dimensional algebras, which are classified by equation \eqn{classbr},
may be realized as subalgebras of ${\cal F}(\R^4)$. Namely we will consider all
 inequivalent cases described by the choice of the
parameters.

Let us consider first the case when the only non zero parameter is $n$. We
have then
\beq
\{x,y\} =ny,~~~~    \{y,w\}=0,~~~~\{w,x\}=-nw.    \label{re3}
\eeq
This is the Lie algebra of $SB(2,\C)$ for any $n \ne 0$.
We can define our map $\pi:\R^4 \rightarrow \R^3$ to be given by
\beq
x=p_{1}+p_{2},~~~~y=e^{nq_{1}},~~~~w=e^{nq_{2}} \label{sa1}
\eeq
or, in terms of complex valued variables,
\beq
x=i(z_{1}+z_{2}),~~~~y=ie^{inz_{1}^{*}}~~~~w=ie^{inz_{2}^{*}}~. \label{sa2}
\eeq
Both the subalgebras \eqn{sa1}, \eqn{sa2}, reproduce relations \eqn{re3}.

A different case corresponds to n=0 and all other parameters are positive
numbers.
In this case we obtain the $SU(2)$  algebra.
A real realization of this algebra is given by
\beq
x=\frac {1}{2}(q_{1}q_{2}+p_{1}p_{2})\sqrt {bc},~~~~
y=\frac {1}{2}(p_{1}q_{2}-q_{1}p_{2})\sqrt {ac},~~~~
w=\frac {1}{4}(p_{1}^2 + q_{1}^2 - p_{2}^2 - q_2^2)\sqrt {ab}
\eeq

while a complex one is
$$
x=\frac{1}{2}(z_1^*z_2^*-z_1 z_2) + \frac{i}{4}(z_1^2-z_2^2-z_1^{*2}+z_2^{*2})
{}~~~~~~~
y=\frac{1}{2}(-z_1^*z_2^*+z_1 z_2) + \frac{i}{4}(z_1^2-z_2^2-z_1^{*2}+z_2^{*2})
$$
$$
w=\frac{1}{2}(z_1^*z_2-z_2^* z_1)
$$

\noindent If each one of the three  parameters $a,b,c$ is negative, a
realization
of the same algebra will be obtained by changing  the sign of all the three
 generators.

If one of the parameters is negative, say $b$, and the  others are positive, we
 have
a classical realization of the $SL(2, \R)$ algebra and a possible choice for
the classical \js~map in terms of real functions is
given by

\beq
x=\frac {1}{2}(p_{1}q_{2}+p_{2}q_{1})\sqrt {-bc},~~~~
y=\frac {1}{2}(p_{1}q_{2}-p_{2}q_{1})\sqrt {ac},~~~~
w=\frac {1}{2}(p_{1}q_{1}-p_{2}q_{2})\sqrt {-ab}
\eeq
 while in terms of complex functions it is
\beq
x=\frac {i}{2}(z_{1}z_{2}^*+z_{2}z_{1}^*)\sqrt {-bc},~~~~
y=\frac {i}{2}(z_{1}z_{2}^*-z_{2}z_{1}^*)\sqrt {ac},~~~~
w=\frac {i}{2}(z_{1}z_{1}^*-z_{2}z_{2}^*)\sqrt {-ab}.
\eeq

We will discuss now the cases when the parameter $a$ is equal to zero and b,
c
are different from zero. If both $b$ and $c$ are positive we
have the three-dimensional $E_{2}$ algebra.
Then, a realization in terms of real phase space
variables is
\beq
x=(p_{1}+p_{2})\sqrt {bc}~~~~
y=\sin (\frac {q_{1}+q_{2}}{2})~~~~
w=-\sqrt {\frac {b}{c}}\cos (\frac {q_{1}+q_{2}}{2})  \label{eucl1}
\eeq
and a complex realization is
\beq
x=i(z_{1}+z_{2})\sqrt {bc}~~~~
y=i\sin (\frac {z_{1}^*+z_{2}^*}{2})~~~~
w=-i\sqrt {\frac {b}{c}}\cos (\frac {z_{1}^*+z_{2}^*}{2})  \label{eucl2}
\eeq

If one coefficient is
positive, say $c$, and the other one, say $b$, is negative, we have the
Poincar\'e algebra in three dimensions. A possible real realization is
\beq
x=(p_{1}+p_{2})\sqrt {-bc},~~~~
y=\sinh (\frac {q_{1}+q_{2}}{2}),~~~~
w=\sqrt {\frac {-b}{c}}\cosh (\frac {q_{1}+q_{2}}{2})
\eeq
and a complex one is
\beq
x=i(z_{1}+z_{2})\sqrt {-bc},~~~~
y=i\sinh (\frac {z_{1}^*+z_{2}^*}{2}),~~~~
w=i\sqrt {\frac {-b}{c}}\cosh (\frac {z_{1}^*+z_{2}^*}{2})
\eeq

The last unexploited case, namely when only one coefficient is
different from
zero, describes the Heisenberg-Weyl group, $H(1)$. By choosing $c$ to be the
non-zero coefficient, we may consider, for example,
the following real realization for the algebra of this group
\beq
x=cq_1~~~~y=p_{1}q_{2}~~~~w=-q_2     \label{heire}
\eeq
and the complex one
\beq
x=icz^*_1~~~~y=iz_{1}z^*_{2}~~~~w=-iz^*_2.      \label{heicom}
\eeq
Of course, there are many other realizations in terms of functions on ~~$\R^4,$
{}~~for
the Heisenberg--Weyl case.

The final case of an abelian algebra, which is obtained by taking all the
coefficients in \eqn{classbr} equal to zero, may also be realized by many
functions.
We give for completeness one of the possible realizations
\beq
x=q_{1}p_{1}~~~~y=q_{2}p_{2}~~~~w=I
\eeq

Now we will discuss the case when $a$ is equal to zero and the other
three parameters
$b,c,n$ are not equal to zero. This case may be reduced to the considered ones
by linear transformation of generators $y$ and $w$ which is equivalent to
diagonalyzing the 2- dimensional matrix
$$
\left(\bea{cc}
    n & c  \\
    -b & n
\ea \right)
$$
the eigenvalues will be
\beq
\lambda_1=n+\sqrt {-bc}~~~~\lambda_2=n-\sqrt {-bc}
\eeq
while the new basis of functions is
$$
X=x
$$
$$
Y=\frac{1}{2} y + \sqrt {-\frac{c}{b}} w
$$
\beq
W=\frac{1}{2} y - \sqrt {-\frac{c}{b}} w
\eeq
and
satisfies the Poisson brackets
\beq
\{X, Y\}=\lambda_1 Y~~~~\{X, W\}=\lambda_2 W~~~~\{Y, W\}=0
\eeq
Then, depending on the values of the two parameters $\lambda_1$, $\lambda_2$,
we may identify one of the previous algebras.

Let us consider some realization of these relations. If the coefficients b
and c have different signs a real valued realization may be chosen to be
$$
x=p_1+p_2
$$
$$
y=e^{\lambda_1 q_1} + e^{\lambda_2 q_2}=
e^{(n+\sqrt{-bc})q_1} + e^{(n-\sqrt{-bc})q_2}
$$
\beq
w=\sqrt{-\frac{b}{c}} (e^{\lambda_1 q_1} - e^{\lambda_2 q_2})=
\sqrt{-\frac{b}{c}} (e^{(n+\sqrt{-bc})q_1} -
e^{(n-\sqrt{-bc})q_2})
\eeq
and a complex one is
$$
x=i(z_1+z_2)
$$
$$
y=ie^{\lambda_1 z_1^*} + e^{\lambda_2 z_2^*}=
ie^{(n+\sqrt{-bc})z^*_1} + e^{(n-\sqrt{-bc})z^*_2}
$$
\beq
w=i\sqrt{-\frac{b}{c}} (e^{\lambda_1 z^*_1} - e^{\lambda_2 z^*(_2})=
\sqrt{-\frac{b}{c}} (e^{(n+\sqrt{-bc})z^*_1} -
e^{(n-\sqrt{-bc})z^*_2})
\eeq

If the coefficients have equal signs we may choose
$$
x=p_1+p_2
$$
$$
y=
e^{(n+\sqrt{bc})q_1} + e^{(n-\sqrt{bc})q_2}
$$
\beq
w=
-\sqrt{\frac{b}{c}} (e^{(n+\sqrt{bc})q_1} -
e^{(n-\sqrt{bc})q_2})
\eeq
in terms of real functions, and
$$
x=i(z_1+z_2)
$$
$$
y=i(e^{(n+\sqrt{bc})z^*_1} + e^{(n-\sqrt{bc})z^*_2})
$$
\beq
w=
-i\sqrt{\frac{b}{c}} (e^{(n+\sqrt{bc})z^*_1} -
e^{(n-\sqrt{bc})z^*_2})
\eeq

If one of the coefficients is equal to zero, for example $c=0$, equations
\eqn{classbr} become
relations
\beq
\{x,y\}=ny,~~~~\{y,w\}=0,~~~~\{x,w\}=nw-by
\eeq
and can be realized by
\beq
x=p_1+p_2~~~~y=e^{n q_2}~~~~w=-b q_1 e^{n q_2}
\eeq
or by
\beq
x=i(z_1+z_2)~~~~y=ie^{n z^*_2}~~~~w=-b iz^*_1 e^{n z^*_2}
\eeq

All the other cases may be reduced to the considered ones.

Thus we have shown by construction that all 3--dimensional Lie algebras can
be realized through a classical \js map.

\section {Deformed \js map}

In the previous section we have realized
all the 3--dimensional Lie algebras as algebras of functions on $\R^3$ and
we have found a map, $\pi$, from $\R^4$ to $\R^3$ (what we call classical \js
map) which gives a realization of 3--dimensional algebras in terms of
functions on $\R^4$. By replacing $z,~z^*$, with creation and annihilation
operators we obtain a realization of these algebras in terms of operators,
modulo ordering problems, which have to be taken into account case by case
when they occur.

Our next goal will be to consider deformed algebras, which are known in the
literature as non commutative Hopf algebras or
quantum groups, and we will look for a realization of the associated
commutator algebras  in terms of
Poisson brackets. We will pose the
problem in the following way:

\noindent Given a 3--dimensional Lie algebra, $\g$, together with its classical
realizations in ${\cal F}(\R^3)$ and in ${\cal F}(\R^4)$, connected by the
classical \js map
$$
\pi : \R^4 \rightarrow \R^3
$$
which we assume to be known now, we will look for the possibility to exhibit
a \js map for the
corresponding deformed algebra $\g_q$, that is to say a realization of $\g_q$
in terms
of functions on $\R^4$ and Poisson brackets. We will approach the problem first
 by looking for a map
$$
\phi : \R^3 \rightarrow \R^3
$$
which gives the deformed generators as functions of the undeformed ones,
then, by using the classical \js map $\pi$ of the
undeformed algebra, we will obtain the wanted map by composing $\phi$ and
$\pi$
$$
\pi '= \phi \circ \pi
$$
This situation may be visualized by the following diagram
$$
\bea{ccc}
      &  &    \\
      &  &\R^3\\
  &{\scriptstyle \pi'}\!\!\nearrow ~~  & ~ \uparrow {\scriptstyle \phi}\\
\R^4 & \stackrel{\pi}{\longrightarrow} & \R^3\\
      &  &     \\
\ea
$$
The map $\pi'$ is what we call the deformed \js map.
Starting with a q-deformed algebra we consider the associated commutator
algebra and
realize it in terms of Poisson brackets looking for a map $\phi$ to be
composed with the classical \js map $\pi$. Of course in some cases one may find
convenient to look for $\pi'$ directly, without bothering with the
factorization. We give a solution,  that is a deformed \js map, for
${\cal {SL}}_q(2, \R)$, ${\cal E}_{q}(2)$ and for ${\cal H}_q(1)$ (${\cal
H}(1)$
 is the Lie algebra of the Heisenberg--Weyl group).

Let us consider the Lie algebra ${\cal {SL}}(2)$ given by relations
\eqn{su2com}. A deformation of this realization is obtained (\cite{bied},
\cite{mcfar}) by setting
\beq
f(\hat n_{i}) = \sqrt \frac {\sinh \lambda \hat n_{i}}{\hat n_{i} \sinh
\lambda}
{}~~~~~\lambda \in \R - \{0\}
\eeq
and defining
$$
X_{q+} = f(\hat n_{1})a_{1}^{\dag}a_{2}f(\hat n_{2})=a_{1q}^{\dag}a_{2q}
$$
$$
X_{q-} = f(\hat n_{2})a_{2}^{\dag}a_{1}f(\hat n_{1})=a_{2q}^{\dag} a_{1q}
$$
\beq
X_{q3} = \frac{1}{2}(\hat n_{1} - \hat n_{2})~,~~~~q=e^{\lambda}
\label{quanal}
\eeq
we get a realization of $SL_{q}(2)$. This is defined to be
$$
[X_{q+}, X_{q3}] = - X_{q+}
$$
$$
[X_{q-}, X_{q3}] = X_{q-}
$$
\beq
[X_{q+}, X_{q-}] = \frac {\sinh 2 \lambda X_{q3}}{\sinh \lambda} . \label{cns}
\eeq

\noindent
This realization is a generalization of the \js  map to ${\cal {SL}}_q(2)$.

Let us look now at the classical counterpart and consider, to be definite,
${\cal {SL}}(2, \R)$. What we mean by classical version of the commutators
 algebra associated with
 ${\cal {SL}}_q(2)$ is a Poisson algebra of functions such that Poisson
brackets have the same form of commutators \eqn{cns}
\beq
\{{\cal X}_{q \pm}, {\cal X}_{q 3}\} = \mp \tilde {\cal X}_{q \pm}
 ~~~~~~~\{{\cal X}_{q +}, {\cal X}_{q -}\} =
\frac {sinh 2 \lambda {\cal X}_3}{sinh \lambda}. \label{defpb}
\eeq
It can be checked that these brackets are in fact Poisson brackets,
evaluating the associated Casimir 1--form, $\alpha_q$, and checking the Jacobi
identity, which, as we said in section 3, is equivalent to show that
$d\alpha_q \wedge \alpha_q=0$. The Casimir 1--form for the Poisson structure
given by relations \eqn{defpb} is
\beq
\alpha_q=d~\{{\cal X}_{q+}{\cal X}_{q-} + \frac{1}{\lambda sinh \lambda}
(sinh\lambda {\cal X}_{3})^2\}
\eeq
verifying
$$
d\alpha_q\wedge\alpha_q=0 \label{alp1}
$$
Let us look now at a realization of this Poisson algebra in terms of
functions of the undeformed algebra. A classical
realization of ${\cal {SL}}(2, \R)$, is given by relations \eqn{al} or  by
their
real version \eqn{real}. We will indicate with the subscript $0$ the
undeformed Poisson structure.

To obtain the wanted  realization in
terms of functions
on $\R^3$ while using the Poisson brackets \eqn{unre}, i. e. to find the map
$\phi:\R^3 \rightarrow \R^3$, we have to solve the
three differential equations
\beq
\{{\cal X}_{q \pm}, {\cal X}_{q 3}\}_0 = \mp {\cal X}_{q \pm}
 ~~~~~~~\{{\cal X}_{q +}, {\cal X}_{q -}\}_0 =
\frac {sinh 2 \lambda {\cal X}_3}{sinh \lambda}. \label{deffpb}
\eeq
It can be checked that a solution is given by
\beq
{\cal X}_{q \pm}=\sqrt{\frac{sh \lambda ({\cal J} + {\cal X}_3)
sh \lambda ({\cal J} - {\cal X}_3)}{({\cal J} + {\cal X}_3)({\cal J} -
{\cal X}_3) sh^2 \lambda}} \sqrt {\frac{sh \lambda}{\lambda}} {\cal
X}_{\pm}~,
{}~~~~~~{\cal X}_{q 3} ={\cal X}_{3}  \label{qal}
\eeq
where
$$
{\cal J}= \sqrt{({\cal X}_+ {\cal X}_- + {\cal X}_3^2)}.
$$
Composing the map $\phi$ with the classical \js map, $\pi$, given by
\eqn{al}, we obtain the wanted map $\pi'$
\beq
\pi'^*({\cal X}_{q +}) =i\zeta_1^* \zeta_2,~~~~~~\pi'^*({\cal X}_{q -}) =
i\zeta_2^*
\zeta_1,~~~~~~\pi'^*({\cal X}_{q 3}) =\pi^*({\cal X}_{3})  \label{qal4}
\eeq
with
\beq
\zeta_i= \tilde f(iz_i z_i^*) z_i~~~~~i \in \{1,2\},    \label{nonlin}
\eeq
\beq
\tilde f (x)= \{ \frac {sinh \lambda}{\lambda} \}^{\frac{1}{4}} f(x)
{}~~~~f(x)=\sqrt{\frac {sh \lambda x}{x sh \lambda}}.  \label{ftl}
\eeq

 The map $\pi'$
is what we call classical \js map for the deformed case.

In comparing the two realizations \eqn{quanal} and \eqn{qal4}, we notice that
in the last one the factor $((sh \lambda) \/ \lambda )$ appears signalling that
in the deformation process also the Dirac map is deformed.
A similar construction can be done with real functions.
Let us introduce new real variables in $\R^4$
\beq
{\pi}_{i}=\tilde {f} (p_{i} q_{i}) p_{i},~~~~~~ {\xi}_{i}=\tilde {f}
(p_{i} q_{i})
q_{i} \label{renonlin}
\eeq
where the function $\tilde f$ is given by relation \eqn{ftl}.
Then the map $\pi'$ is given by:
\beq
{\cal Y}_{q +}=\pi_{1}\xi_{2},~~~~~~~{\cal Y}_{q -}=\pi_{2}\xi_{1},~~~~~~~~
{\cal Y}_{q 3}={\cal Y}^3
\eeq
One can check that the Poisson algebra so obtained is again the q--deformed
algebra ${\cal {SL}}(2, \R)$ defined by relations \eqn{cns}.

Let us now consider the euclidean algebra ${\cal {E}}(2)$. The Lie algebra
commutation
relations are
\beq
[P_x, P_y] = 0,~~~~,[J, P_x] = P_y,~~~~[J, P_y] = -P_x.
\eeq
A deformation of this algebra which has been shown to be a non commutative Hopf
algebra has been given in \cite{cel}, the commutation relations being
\beq
[P_{x q}, P_{y q}] = 0,~~~~,[J_q, P_{x q}] = P_{q y},~~~~[J_q, P_{y q}] =
-\frac {1} {\lambda} sh (\lambda P_x). \label{chnsa}
\eeq
Also in this case it can be checked that the classical analogue of
\eqn{chnsa}
\beq
\{ {\cal P}_{x q}, {\cal P}_{y q} \}=0,~~~~
\{ {\cal J}_q, {\cal P}_{x q} \}= {\cal P}_{y q},~~~~~
\{ {\cal J}_q, {\cal P}_{y q} \}=-\frac{1}{\lambda} sh (\lambda{\cal P}_{x}).
\eeq
satisfies the Jacobi identity, by evaluating the Casimir 1--form $\alpha_q$
and by checking that $d\alpha_q \wedge \alpha_q=0$. As it can be verified the
1--form is
\beq
\alpha_q=\frac{1}{2}d~\{{\cal
P}_{y}^2+\frac{2}{\lambda^2}(sinh\frac{\lambda}{2}
{\cal P}_{x})^2\}
     \label{alp2}
\eeq
and it satisfies the wanted equation.
For the undeformed algebra a classical \js map is given by relations
\eqn{eucl2}, where we make the following identification
$$
{\cal P}_x = y,~~~~~{\cal P}_y=w,~~~~~{\cal J}=x.
$$
 To find the map $\phi$ such that
$$
{\cal P}_{x q}=\phi^*({\cal P}_x),~~~~
{\cal P}_{y q}=\phi^*({\cal P}_y),~~~~
{\cal J}_q=\phi^*({\cal J})
$$
we have to solve the three differential equations
\beq
\{ {\cal P}_{x q}, {\cal P}_{y q} \}_0=0,~~~~
\{ {\cal J}_q, {\cal P}_{x q} \}_0= {\cal P}_{y q},~~~~~
\{ {\cal J}_q, {\cal P}_{y q} \}_0=-\frac{1}{\lambda}
sh (\lambda{\cal P}_{x q}).
\eeq
It can be verified that a solution is
\beq
{\cal P}_{x q}={\cal P}_x,~~~~~~
{\cal P}_{yq}=\sqrt {-\frac{2}{\lambda^2} ch (\lambda {\cal P}_x)},~~~~
{\cal J}_q=\frac {{\cal J}}{{\cal P}_y} \sqrt {-\frac{2}{\lambda^2}
ch (\lambda {\cal P}_x)}.
\eeq
By composing the map $\phi$ with the classical \js map \eqn{eucl1} or
\eqn{eucl2} we obtain
the \js map for the deformed algebra ${\cal E}_{q}(2)$
\beq
\pi'^*{\cal P}_{xq}=\pi^* {\cal P}_x,~~~~~
\pi'^*{\cal P}_{yq}=\sqrt {-\frac{2}{\lambda^2} ch (\lambda \pi^*{\cal
P}_x)},~~~
\pi'^*{\cal J}_q=\frac {\pi^* {\cal J}}{\pi {\cal P}_y} \sqrt {-\frac{2}
{\lambda^2} ch (\lambda \pi^*{\cal P}_x)}.
\eeq

Finally we consider the Heisenberg--Weyl group $H(1)$. The Lie brackets for the
 algebra ${\cal H}(1)$ are given by
$$
[X, Y]= W,~~~~~[W,X]=0,~~~~~[W, Y]=0.
$$
A realization of this algebra as a Poisson subalgebra of ${\cal F}(\R^4)$ is
given by relations \eqn{heire} and \eqn{heicom}, which are respectively a real
and a complex realization. A q--deformation of the Heisenberg--Weyl algebra,
which turns out to be a non commutative Hopf algebra has been given in
\cite{cel}, the deformed commutators being
\beq
[X_q, Y_q]= \frac{sh \lambda W}{\lambda},~~~~~[W_q,X_q]=0,~~~~~[W_q, Y_q]=0.
\label{ncf}
\eeq
The classical analogue of \eqn{ncf}, that is
\beq
\{{\cal X}_q, {\cal Y}_q\}= \frac{sh \lambda {\cal W}}{\lambda},~~~~~\{{\cal
W}_q,
{\cal X}_q\}=0,~~~~~\{{\cal W}_q, {\cal Y}_q\}=0
\eeq
 can be shown to be a Poisson algebra by evaluating the associated Casimir
1--form, which turns out to be
\beq
\alpha_q=d \frac{cosh (\lambda {\cal W})}{\lambda^2}             \label{alp3}
\eeq
and verifying that
$d\alpha_q \wedge \alpha_q=0$.
A realization of this algebra as a Poisson subalgebra of ${\cal F}(\R^3)$,
whose generators are functions of the undeformed ones
$$
{\cal X}_{iq} = \phi^*({\cal X}_{i})
$$
can be verified to be
\beq
{\cal X}_{q}={\cal X},~~~~~
{\cal Y}_{q}=\frac{sh (\lambda {\cal W})}{\lambda} \frac {{\cal Y}}{{\cal
W}},~~~~~{\cal W}_{q}=\frac {sh (\lambda {\cal W})}{\lambda}.
\eeq
Composing then the map $\phi$ with the map $\pi$ (the classical \js map),
we obtain
\beq
\pi'^*({\cal X}_q) = iz_1,~~~~~\pi'^*({\cal Y}_q) = iz^*_1 \frac{sh (\lambda
z_2)}{\lambda},~~~~~\pi'^*({\cal W}_q) = i\frac{sh (-\lambda z_2)}{\lambda}
\label{jswei}
\eeq
where we have used the complex realization of $\pi$ given by relations
\eqn{heicom}. Relations\eqn{jswei} give then a possible choice for the
classical deformed \js map of the Heisenberg--Weyl algebra.

Before closing this section we would like to make a few remarks on the
deformation procedure regarded in a classical setting.

As we have briefly
sketched in section 3, Poisson structures on $\R^3$ can be characterized by
their Casimir 1--form, $\alpha$, satisfying the condition ~~$d\alpha \wedge
\alpha=0$~~ and their properties can be restated in terms of the 1--form.
The most general expression for $\alpha$ is given by relation \eqn{alph},
which we write again for our convenience
$$
\alpha = n(y dw-w dy) + d \frac{1}{2} (ax^2+by^2+cw^2).
$$
If we consider now
\beq
\alpha = n(f~dg -g~df)+ d \frac{1}{2} (ah^2+bf^2+cg^2)   \label{defalph}
\eeq
with~~ $f=f(x, y, w, \eta),~~ g=g(x, y, w, \lambda),~~ h=h(x, y, w,
\tau),~~$
then \eqn{defalph} provides a general deformation of our algebras as long as
$f(x, y, w, 0)=y,~~ g(x, y, w, 0)=w,~~ h(x, y, w,
0)=x.~~$ If $~~df \wedge dg \wedge dh \ne 0,~~$ we can think of our
deformation as a change of coordinates on $\R^3$ associated with possible
change of coordinates on $\R^4$.
We can say
that we are picking a different subalgebra of functions in the Poisson
algebra $({\cal F}(M), \{~,~\}_0)$, or we can say that we have performed a
nonlinear noncanonical transformation taking us from one Poisson bracket to
another. Both view points are equivalent and acceptable.
Let us consider now our specific examples. A possible
multi-parametric deformation of the
1--form $\alpha$ would be
\beq
\alpha(\lambda, \eta, \tau) = n([y]_\eta d[w]_\lambda - [w]_\lambda d[y]_\eta)
+ d \frac{1}{2} (a[x]_\tau^2+b[y]_\eta^2+c[w]_\lambda^2)
\eeq
where $[r]_\sigma$ are the Tchebyshev polynomials
$$
[r]_\sigma=\frac{\sigma^r-\sigma^{-r}}{\sigma - \sigma^{-1}}.
$$
We find then that ${\cal S}{\cal L}(2, \R)$ algebra is obtained by
setting $n=0$ and choosing $\tau \rightarrow 1, \eta \rightarrow 1$, we get
the deformed 1--form for ${\cal S}{\cal L}(2, \R)$ \eqn{alp1}, which leads to
the
deformed Poisson brackets \eqn{defpb}. For $n=0, c=0$ and $\tau \rightarrow
1$ we get the deformed 1--form \eqn{alp2} and then ${\cal E}_q(2)$; finally,
for $n=0, b=0$ and $\tau \rightarrow 1$,
we get the 1--form \eqn{alp3} which leads to ${\cal H}_q(1)$.

Despite of the fact that we have been able to obtain some known results, up to
now this only provides a way to reinterpret the deformation procedure at the
classical level, not to give new Hopf algebras, the reason being that we
are not able to give a prescription which should be satisfied by the deformed
Poisson algebras to make the corresponding operator algebras into Hopf
algebras. We hope to come back to this issue in the near future.

\section{Analogue of quantum and classical \js map for arbitrary Lie algebras}

We make a few comments here on the possibility to realize commutation relations
 for a Lie algebra of higher dimensions, in terms of Poisson brackets.
Realizations for an arbitrary Lie algebra are based
on a known procedure from second quantization (\cite{ber}).
To exemplify, let us consider three $N \times N$ matrices $A,~B,~C$ which
satisfy the relation
$$[A,B] =C;$$
we can construct then three operators
$$
\hat A=A_{ik}a_i^{\dag}a_k~~~~
\hat B=B_{ik}a_i^{\dag}a_k~~~~
\hat C=C_{ik}a_i^{\dag}a_k
$$
which reproduce the initial commutation relation
$$
[\hat A,\hat B] =\hat C.
$$
Here operators $a_i,~a_i^{\dag}$ are either bosonic ones i.e.
$$
[a_i,a_k^{\dag}]=\delta_{ik}~~~~~~~~~~~~~[a_i,a_k]=0
$$
or fermionic ones, i.e.
$$
[a_i,a_k^{\dag}]_+=\delta_{ik}~~~~~~~~~~~~~[a_i,a_k]_+=0
$$
Due to that, given M generators-matrices
of any N--dimensional Lie algebra representation
$(L_\alpha)_{ik},~~~~ i,k=1,...N~~~~~\alpha=1,...M$ with
\beq
[L_\alpha ,L_\beta ]=c_{\alpha \beta}^\gamma L_\gamma
\label{2comm}
\eeq
then the operators
\beq
\hat L_\alpha = (L_\alpha)_{ik} a_i^{\dag} a_k \label{op}
\eeq
realize the same Lie algebra in both cases of bosonic
and fermionic nature of the operators $a_i, a_i^{\dag}$. Below we will have in
mind bosonic operators. Then it is easy to check that classical analogues of
\eqn{op}
\beq
\tilde L_\alpha=(L_\alpha)_{ik} z_i^*z_k
\eeq
give a realization of the same Lie
algebra in terms of Poisson brackets. Namely we have
\beq
\{\tilde L_\alpha, \tilde L_\beta \}=c_{\alpha \beta}^{\gamma } \tilde L_
\gamma~.
\eeq
In this sense we could always have Lie algebra realizations
in terms of Poisson brackets of this type. Once we have this realization for
Lie
algebras, we can extend it to deformed algebras.

We can cast these comments in our geometrical framework by noticing that
with any $N \times N$ matrix A we can associate a vector field $X_A=A^i_j
x^j \frac{\del}{\del x^i}$ acting on $\R^N$. By considering now $T^* \R^N$ we
get a natural realization in terms of Poisson brackets by using functions
$f_A=A^i_j x^j p_i$. If, on the other hand, we complexify $\R^N$ to get
$\C^N$ we can consider a natural symplectic structure $\omega = \sum_a
dz_a^{\dagger} \wedge dz_a~~~~~a\in \{1,...N\}$ and associate a complex
valued function with any matrix, namely $\zeta_A=A^i_jz^jz_i^*$. It
would be possible to consider also Grassmann variables to get the classical
analogue of fermionic variables.

It should be remarked that once we realize our algebra in terms of vector
fields, we can perform any non linear coordinate transformation on $\R^N$
without spoiling the commutation relations of our vector fields.

{}From what we have said, it is clear that out of any finite dimensional
representation of a Lie algebra we can construct a realization in terms of
Poisson brackets and use this realization to undertake deformations.

\section {Extension of Jordan-Schwinger map to $U_{q}(n)$}.
We will consider  now the Lie algebra of $U (n)$ and its deformation, which is
given in \cite{jim} and, in terms of q--oscillators, by \cite{sun} ; we will
look for an extension of previous results to it.
The generators $T_{ik}$ of the Lie algebra $\cal U (n)$ may be
constructed  from the creation and annihilation operators $(a_{i},
a_{i}^{\dag}) (i=1,..,n)$ in such a manner
 \beq
 T_{ik} = a_{i}^{\dag}a_{k}   \label{gen}
 \eeq
and they will satisfy the commutation relations
 \beq
 [T_{ik}, T_{mn}] = T_{in}\delta _{km} - T_{mk}\delta_{in}
 \eeq
A q-deformation of this algebra may be produced
using the relations
$$
a_{iq}=a_i f(\hat n_i)
$$
$$
 T_{qik} = \frac {1}{\sinh \lambda}T_{ik}\sqrt {\frac{\sinh \lambda(T_{ii}+1)
 \sinh \lambda T_{kk}}{(T_{ii}+1)T_{kk}}}=a_{iq}^{\dag} a_{kq}~~~~i<k
$$
\beq
T_{qii} = T_{ii}       \label{genq}
 \eeq
 with
 \beq
 T_{qik} = T_{qki}^{\dag}~~~~  for~~ k<i
 \eeq
and inverse
 $$
 T_{ik} = (\sinh \lambda)~ T_{qik} \sqrt {\frac {T_{qkk}(T_{qii}+1)}{\sinh
 (\lambda T_{qkk})\sinh (\lambda T_{qii} + 1)}}~~~~i<k
$$
$$
 T_{ik} = T_{ki}^{\dag}~~~~  k<i
 $$
\beq
T_{ii} = T_{qii}~.
\eeq
 The operators $T_{qik}$ satisfy the following relations
 \beq
 [T_{qik}, T_{qmn}] = 0 ,  i\neq m,n~~~~ and~~~ k\neq m,n
\eeq
 Since the relations \eqn{gen} hold, the generators $T_{qik}$ given by
\eqn{genq} may be realized as functions of creation and annihilation operators.
Let us specialize to the case N=3. The deformed $\cal U^q (3)$ has three
subalgebras
$\cal U^q (2)$ which are realized by
\beq
T_{ii}^q,~~T_{ik}^q,~~T_{ik}^{q \dag},~~T_{kk}^q~~~~~i,k=1,...N
\label{ngen}
\eeq
and happen to be the standard q--deformed $\cal U^q (2)$, with commutation
relations
\beq
[T_{ik}^q,T_{is}^{q \dag}]=-T_{sk}^q F(T_{ii}^q) \label{commun}~.
\eeq
The function $F(x)$ is given by
\beq
F(T_{ii}^q)=\frac{sh \lambda (T_{ii}^q +1)-sh \lambda T_{ii}^q}{sh\lambda}.
\eeq
For $\lambda \rightarrow 0$ this function becomes
\beq
F_l(T_{ii}^q)=\frac{ch \lambda T_{ii}^q} {sh\lambda}. \label{fun}
\eeq
This expression will be useful in the classical analogy which we are going
to treat.

Let us consider now a classical realization of the q--deformed $\cal U (n)$
algebra in
terms of Poisson brackets. The classical \js map gives
$$
{\cal T}^q_{ik}=i\zeta^*_{i}\zeta_{k}
$$
where
$$
\zeta_i=\tilde f(in_i) z_i~,
$$
$\tilde f$ is given by \eqn{ftl}, and
\beq
\{\zeta_{iq},\zeta^*_{kq} \}= \frac {-i \lambda}{sh \lambda}
ch(\lambda |\zeta_i|^2) \delta_{ik}~. \label{last}
\eeq
As we can see from equations \eqn{commun}, \eqn{fun} and \eqn{last},
the realization in terms of Poisson brackets agrees with  the one in
terms of commutators only in the limit $\lambda\rightarrow 0$ or, which is
the same, $q \rightarrow 1$.
Namely we have
\beq
\{{\cal T}_{ik}^q,{\cal T}_{is}^{q *} \}=-{\cal T}_{sk}^q {\cal F}_l({\cal
T}_{ii}^q) + O(\lambda)~.
\eeq
Thus, the q--deformation of the classical $\cal U (3)$ algebra with generators
\beq
{\cal T}_{ik}=i\zeta^*_{i}\zeta_{k}
\eeq
and commutation relations
\beq
\{{\cal T}_{ik},{\cal T}_{is} \}=({\cal T}_{in} \delta_{km} - {\cal
T}_{mk}\delta_{in})
\eeq
gives back the subalgebras $\cal U^q (2)$ in the limit of $\lambda$ small.

\section{\bf Conclusions}
In this paper we have generalized the definition of \js map to any three--
dimensional Lie algebra. Lie algebras are classically realized, in the sense
that the Lie product is given by a Poisson bracket and the generators are
realized as functions. We have considered then examples of deformations of
these Poisson subalgebras, which give, through the Dirac map, non
commutative Hopf algebras already known in the literature. In this classical
setting we reinterpret the
deformation procedure as a deformation of the Casimir 1--form associated to
the Poisson structure characterizing the algebra. It is still missing
 how to implement in our
classical procedure additional ingredients to be related to
the notion of Hopf algebra, which appears once the
algebras are realized in terms of operators. We hope to
come on that in the next future.
All this would allow thinking that new ways leading to quantum groups might
exist starting at this classical level.

There is a second problem which is interesting in our opinion, but still
open. We could think of extending the domain of the Jordan-Schwinger map
to symplectic manifolds different from $\C^n$. As we mentioned above, the given
examples of Lie algebras realizations in
terms of Poisson subalgebras might be related to the $T^*G$ and $T^*(G/K)$
structures. Some other examples of classical realizations of three
dimensional Lie algebras already exist in the literature (\cite{mank}).
Nevertheless the problem of an exhaustive classification of all possible
symplectic realizations of Lie algebras needs extra investigation and we are
working on that.

\section{Aknowledgements}
One of the authors (V. I. Manko) thanks I.N.F.N. and the University of
Napoli for the hospitality.
%
%
%
%

\end{document}